# Visible light-assisted peroxymonosulfate activation by high-purity FeS$_2$ nanoplates for dye pollutant control

Yizhe Huang,[#,a] Yuwen Chen,[#,a] Ke Zhu,[a] Pengfei Li,[b] Xu Wu,[b,*] Rafael Luque,[c,*], Kai Yan[a,d,*]

[a] *Guangdong Provincial Key Laboratory of Environmental Pollution Control and Remediation Technology, School of Environmental Science and Engineering, Sun Yat-sen University, Guangzhou, 510275, China*

[b] *College of Chemistry and Chemical Engineering, Taiyuan University of Technology, Taiyuan 030024, China*

[c] *Universidad ECOTEC, Km 13.5 Samborondón, Samborondón EC092302, Ecuador*

[d] *Guangdong Laboratory for Lingnan Modern Agriculture, South China Agricultural University, Guangzhou 510642, China*

[#] *These authors contributed equally to this work.*

[*] *Corresponding author Email addresses: wuxu@tyut.edu.cn (X. Wu), rluque@ecotec.edu.ec (R. Luque), yank9@mail.sysu.edu.cn (K. Yan)*

**Abstract:** With the rapid industrial development, many dye pollutants have entered the water, along with heavy metals like As, leading to complex pollution that threatens the ecological environment and human health. Therefore, designing an effective strategy for treating complex dye wastewater is urgent. Herein, we have constructed high-purity pyrite FeS$_2$ nanoplates as bifunctional catalysts for the simultaneous removal of dyes and arsenite (As(III)). We have developed a visible light-assisted Fenton-like system based on peroxymonosulfate (PMS) to remove typical organic dye pollutants (e.g., rhodamine B, malachite green, methyl orange, methylene blue) and toxic As(III). The FeS$_2$/Light/PMS system can remove 100% of all dye pollutants and more than 85% of As(III) within 18 min simultaneously with high cyclic stability, which is believed to be effective in complex pollution control. It is confirmed that Fe is the main active center, and the free radical and non-free radical pathways synergistically enhance the oxidation of dye pollutants, in which $SO_4^{\cdot-}$ and $^1O_2$ are the main reactive oxygen species. This work presents a facile strategy for developing a metal sulfide-based Fenton-like system for treating complex dye wastewater.

**1. Introduction**

Textile wastewater is heavily laden with dye pollutants, which are continuously released into the aquatic environment [1,2]. Moreover, the existence of heavy metals (e.g., Cr, Pb, and As) in textile wastewater exacerbates the toxicity of effluent, posing a threat to the ecological balance in water [3]. However, dye pollutants are difficult to biodegrade naturally, and traditional wastewater treatment technologies lack efficient and thorough means to remove them [4], accumulating organic dyes in the environment. Therefore, designing an effective and low-cost method for complex dye wastewater control is urgent.

Advanced oxidation technologies are one of the most effective approaches for dye wastewater treatment, especially Fenton or Fenton-like oxidation based on peroxymonosulfate (PMS) [5,6]. Homogeneous Fenton reaction utilized $Fe^{2+}$ catalyst to activate PMS, producing hydroxyl radicals (·OH), sulfate radicals ($SO_4^{\cdot-}$), and other reactive oxygen species (ROS) with strong oxidative ability, which can effectively degrade organic pollutants (e.g., dyes). However, this homogeneous catalytic process is greatly limited by its narrow pH working range of 3.0-4.0 [7,8], and the leaching of $Fe^{2+}$ often leads to excessive iron sludge. In this regard, heterogeneous catalysts can overcome the above issues with great adaptability and reusability. A large number of studies [9–12] thus focus on developing transition metal oxides as Fenton-like catalysts, among them, metal sulfides have shown intriguing capability for Fenton-like reactions [13–16]. Iron sulfides, for example, are commonly found in natural ores, with pyrite $FeS_2$ being the most common on the Earth [17]. The inherently abundant $Fe^{2+}$ species in iron sulfides serve as active centers for PMS activation, and further, pyrite $FeS_2$ possesses a more stable structure that can inhibit the dissolving of $Fe^{2+}$. Besides, Zhou et al. [18] reported that $S_2^{2-}$ in $FeS_2$ act as the crucial mediator in the $Fe^{2+}/Fe^{3+}$ redox cycle, maintaining the content of $Fe^{2+}$ and PMS activation efficiency. On the other hand, pyrite $FeS_2$ is also a semiconductor with a narrow bandgap of 0.95-1.00 eV [19,20], which makes it a potential candidate in photocatalysis. Albeit great progress has been made, the exposure of active sites in $FeS_2$ with high purity is still full of challenges. In addition, previous studies mainly concentrate on controlling a single type of pollutant, highlighting the necessity to investigate the control of complex pollutants.

Herein, high-purity pyrite $FeS_2$ nanoplates were constructed by a facile vacuum annealing method for the simultaneous removal of dyes and As(III). Characterized by multiple techniques, the pure pyrite phase and unique nanoplate structure of as-prepared $FeS_2$ catalysts were revealed, which largely expose active sites for pollutant adsorption and PMS activation. Under visible irradiation, $FeS_2$ nanoplates could efficiently activate PMS, removing 100% of 20 mg L$^{-1}$ RhB and 97.2% of 100 μM As(III) within 18 min. The $FeS_2$/Light/PMS system also showed great applicability in degrading other dye pollutants, including MB, MG and MO. Quenching experiments and electron paramagnetic resonance (EPR) were employed to ascertain the main ROS and X-ray photoelectron spectroscopy (XPS) was performed to identify the role of Fe and S species in the reactions. Moreover, the $FeS_2$/Light/PMS system also showed high stability with excellent performance in cycle experiments and low iron leaching. This work provides a facile and promising tactic for the practical treatment of complex dye wastewater.

## 2. Experimental section

*2.1. Chemicals*

Fe powder (Fe, ≥ 99%, Macklin), high purity sulfur powder (S, 99.5%, Macklin), rhodamine B (RhB, $C_{22}H_{24}O_9N_2$, ≥ 98%, Macklin), methylene blue (MB, $C_{22}H_{24}N_2O_8$, 98%, Aladdin), malachite green (MG, $C_{22}H_{23}ClN_2O_8$·HCl, USP, Macklin), methyl orange (MO, $C_{14}H_{14}N_3SO_3Na$, ≥ 99.99%, Macklin), sodium arsenite (NaAsO$_2$, ≥ 90%, Sigma-Aldrich), peroxymonosulfate ( > 95%, Sigma-Aldrich), tert-Butanol (TBA, $C_4H_{10}O$, 99.5%, Alfa Aesar), 1, 4-Benzoquinone (PBQ, $C_6H_4O_2$, 99%, Macklin), methanol (MeOH, $CH_3OH$, ≥ 99.9%, Macklin), 2, 2, 6, 6-Tetramethylpiperidine (TEMP, $C_9H_{19}N$, 97%, Macklin), 5, 5-Dimethyl-1-pyrrolineN-oxide (DMPO, $C_6H_{11}NO$, ≥ 98.0%, Sigma-Aldrich), ethanol ($C_2H_6O$, ≥ 99.5%, Aladdin).

*2.2. Preparation of FeS$_2$ nanoplates*

High-purity pyrite FeS$_2$ nanoplates were fabricated through a facile vacuum annealing method. First, Fe powder and S powder (molar ratio of 1:2.05) with a total mass of 150 mg were weighed, ground, and mixed. The ground mixture was then sealed in a vacuum quartz tube and calcined at 400 °C for 12 h with a ramping rate of 5 °C min$^{-1}$. Cooling down to room temperature, the sample was then completely ground to obtain the FeS$_2$ nanoplates. Detailed characterizations were illustrated in SI (Text S1).

*2.3. Degradation experiments*

RhB and As(III) were exploited to investigate the activity of the prepared catalysts. In a typical trial, 10 mg of FeS$_2$ was added to a reactor containing 50 mL of RhB and As(III) mixed solution, and dark adsorption was performed for 60 min. Then 1.2 mM PMS was added and illuminated under visible light to trigger the photocatalytic reaction. During the process, 0.5 mL of the reacted sample was extracted at 3-minute intervals and immediately mixed with 0.5 mL of methanol to terminate the reaction. The concentration of RhB was measured using an ultraviolet-visible spectrophotometer (UV-Vis, UV-2500, Shimadzu), and that of As species were identified using high-performance liquid chromatography combined with inductively coupled plasma-mass spectrometry (HPLC-ICP-MS). More detailed analyses were depicted in SI (Text S2).

**3. Results and discussion**

*3.1. Characterization of catalysts*

The synthetic diagram of FeS$_2$ is shown in Fig. 1a. X-ray diffraction analysis (XRD) was used to investigate the crystal structure of prepared catalysts as performed in Fig. 1b. According to XRD patterns, the broad peak positioned at 21° could be indexed to the amorphous silica in the substrate, and the as-prepared FeS$_2$ was consistent with the standard card (FeS$_2$, pyrite, #79-0617), indicating the formation of a high-purity pyrite FeS$_2$ without any impurity [21,22]. Furthermore, XPS analysis revealed the unique spectra of Fe 2*p* and S 2*p* on the FeS$_2$ surface. As shown in Fig. 1c, the signal peak at 707.5 eV and 709.2 eV were both ascribed to Fe(II), whilst the latter was considered evidence of cleavage of the Fe-S bond, leading to electron redistribution [23]. Besides, the signal peak at 711.5 eV was ascertained as Fe (III) due to inevitable Fe(II) oxidation on the surface [24]. In the S 2p spectrum (Fig. 1d), the signal peaks at 162.9 eV, 164.1 eV, 165.3 eV and 169.2 eV corresponding to $S_2^{2-}$, $S_n^{2-}$, $S_0$ and $SO_4^{2-}$, respectively, indicating the existence of various sulfur species in $FeS_2$. The above characterization analysis proved that high-purity pyrite $FeS_2$ without other impurities was successfully prepared by the vacuum annealing method. Moreover, the abundant sulfur species promoted the redox cycle of $Fe^{2+}/Fe^{3+}$ in the reaction process and maintained the stability of its metal-active sites in the redox reaction [25].

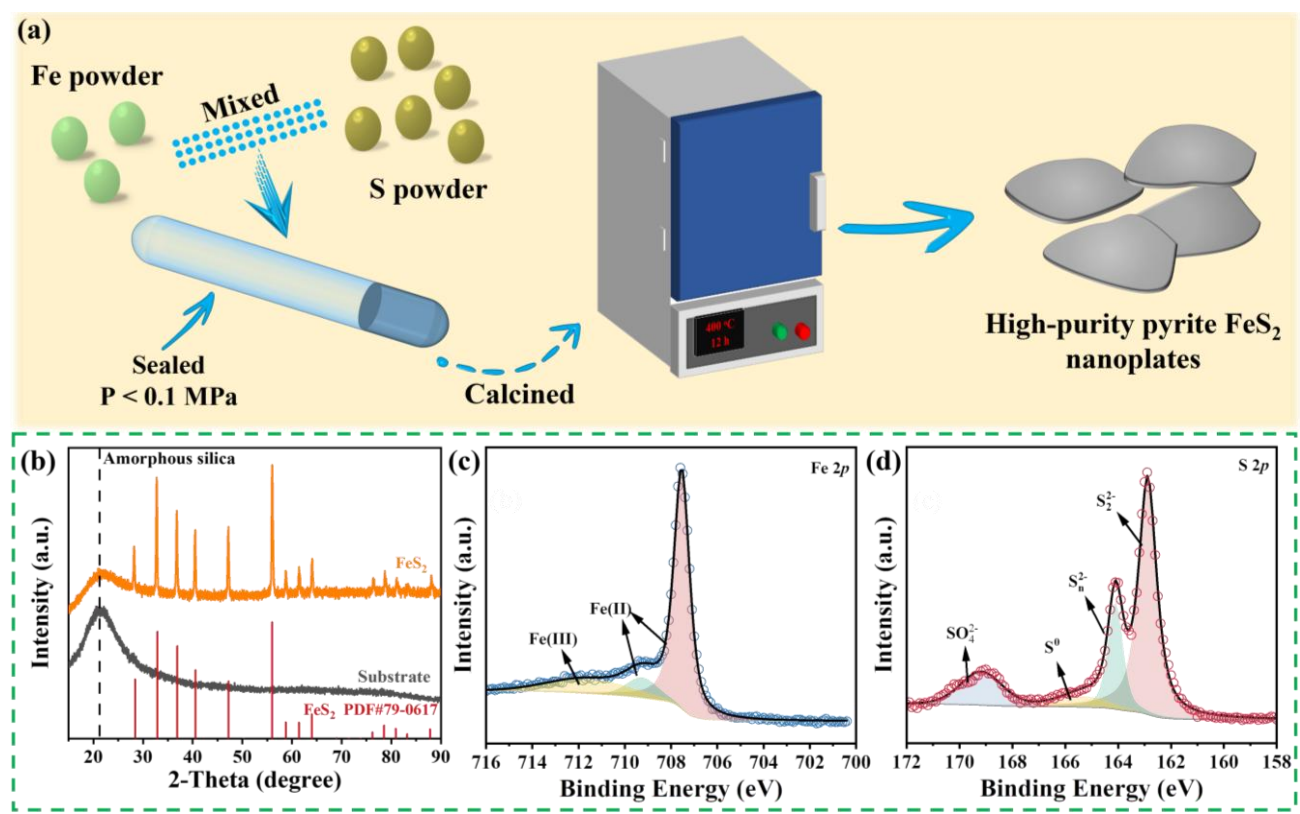

**Fig. 1.** The (a) synthetic schematic diagram, (b) XRD diagram, and XPS spectra of (c) Fe 2*p*, (d) S 2*p* of the high-purity pyrite $FeS_2$.

Transmission electron microscopy (TEM) and high-resolution transmission electron microscopy (HRTEM) were applied to understand the morphology and microstructure of $FeS_2$. TEM images (Fig. 2a-b) revealed that the synthesized pyrite $FeS_2$ has a nanoplate structure, which largely exposes more reactive centers for pollutant adsorption and PMS activation. In the HRTEM image (Fig. 2c), the lattice spacings are measured to be 0.221 nm and 0.243 nm, attributed to the (211) and (210) facets of $FeS_2$, respectively. Moreover, SAED analysis (Fig. 2d) further proved the successful synthesis of $FeS_2$ at the microscopic level.

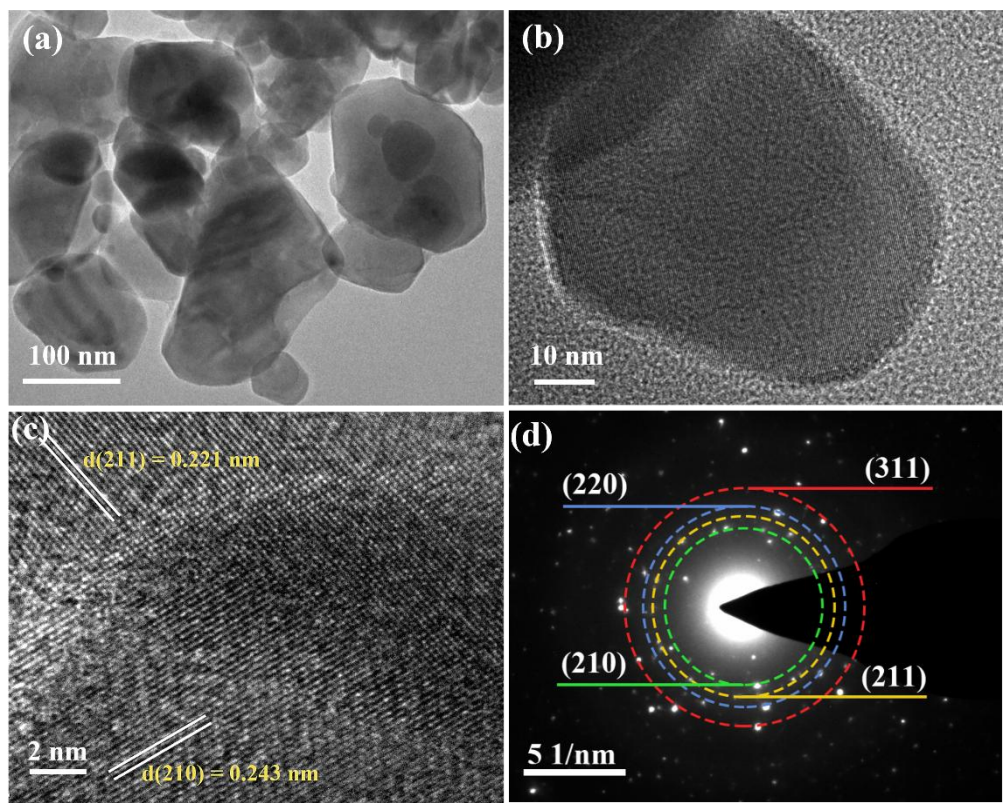

**Fig. 2.** (a-b) TEM images, (c) HRTEM images and (d) SAED pattern of FeS$_2$ nanoplates.

*3.2. Catalytic performance of catalysts*

To estimate the activity of the as-prepared FeS$_2$ nanoplates, degradation experiments were first conducted with RhB as the target pollutant. As shown in Fig. 3a, only the PMS or light systems could remove less than 5% of RhB, indicating the negligible degradation of self-activation of PMS and the visible light alone. Moreover, the removal effect of Light/PMS on RhB was just near 15%. As pyrite FeS$_2$ is a semiconductor material that can utilize visible light, the FeS$_2$/Light system could remove 37.1% of RhB within 18 min. At the same time, it could be observed that the FeS$_2$/PMS system could remove 85.1% of RhB within 18 min, which proved that the FeS$_2$ catalyst has high catalytic activity for activating PMS. Compared with the above reaction systems, the FeS$_2$/Light/PMS system could remove 100% of RhB in 12 min, which showed superior catalytic activity. As shown in Fig. 3b-c, the kinetic rate of the FeS$_2$/Light/PMS system was evaluated to be 0.3629 min$^{-1}$, which is 11.8 times and 2.6 times higher than that of the FeS$_2$/Light system (0.0308 min$^{-1}$) and FeS$_2$/PMS system (0.1376 min$^{-1}$), respectively, suggesting that FeS$_2$ can activate

PMS more efficiently and produce more ROS with the assistance of visible light. Fig. 3d shows the change in the UV-Vis absorbance spectrum of RhB in the FeS$_2$/Light/PMS system. In addition, the degradation efficiencies of different dye pollutants (RhB, MB, MG and MO) in the FeS$_2$/Light/PMS system could all reach 100% within 18 min (Fig. 3e), and the respective TOC removal efficiencies (Table S1) reached 49.2%, 44.6%, 52.1% and 43.8%, respectively. The chromaticity of the dye pollutant solution is effectively removed as it turned colorless and transparent (Fig. 3f), indicating the promising potential of the FeS$_2$/Light/PMS system in dye pollution control.

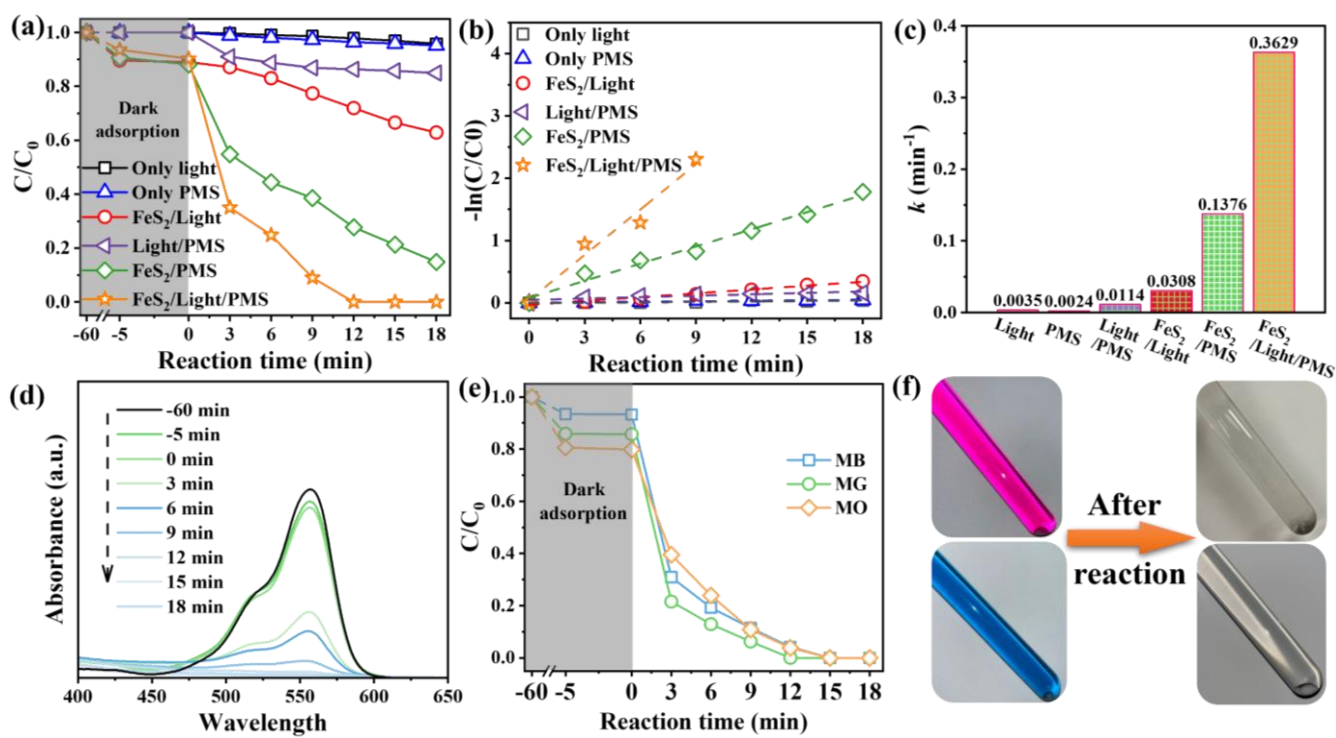

**Fig. 3.** (a) Degradation performance of different reaction systems on ChB and (b, c) corresponding reaction rate constants, (d) UV-vis absorption spectrum changes of RhB in the FeS$_2$/Light/PMS system, (e) degradation effect of different dye pollutants by FeS$_2$/Light/PMS system and (f) color changes of RhB and MB before and after the reaction. (Reaction conditions: [PMS] = 1.2 mM, [Catalyst] = 0.2 g L$^{-1}$, [Pollutants]= 20 mg L$^{-1}$)

To evaluate the catalytic stability of the FeS$_2$/Light/PMS system, cyclic experiments were conducted and the spent FeS$_2$ was characterized by XRD. Even after four cycles of experiments (Fig. 4a), the FeS$_2$/Light/PMS system could still maintain more than 80% of RhB removal efficiency. Besides, after the cyclic reaction, the XRD characterization results (Fig. 4b) prove that the crystal structure of $FeS_2$ has no noticeable change, indicating that $FeS_2$ has good cycle stability. Furthermore, the leaching Fe ions in the $FeS_2$/Light/PMS system during various reaction time were detected by ICP-OES analysis (Table S2). After adding PMS, the leaching concentration of Fe ions reached 1.96 mg $L^{-1}$ in the first 3 min owing to the change in pH of the solution. After 18 min, the final leaching concentration of ions was 2.55 mg $L^{-1}$, indicating that the catalytic active center of $FeS_2$ was stable.

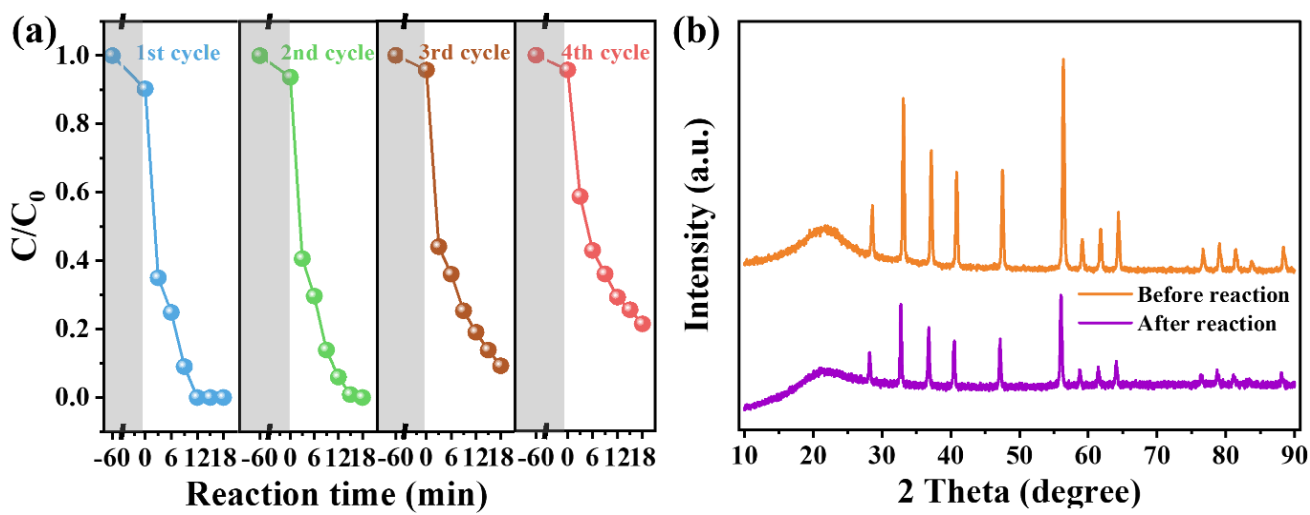

**Fig. 4.** (a) Cycle experiment and (b) XRD of $FeS_2$ before and after the reaction. (Reaction conditions: [PMS] = 1.2 mM, [Catalyst] = 0.2 g $L^{-1}$, [RhB]= 20 mg $L^{-1}$)

Dye wastewater is a complex mixture containing a wide range of dyestuff organic pollutants and heavy metals such as highly toxic As(III). To evaluate the application potential of the $FeS_2$/Light/PMS system in treating complex dye wastewater, the catalytic performance for the synergistic removal of RhB and As(III) was studied, as shown in Fig. 5. When RhB concentration in the solution retained at 20 mg $L^{-1}$ and the concentration of As(III) increased from 50 μM to 200 μM (Fig. 5a-c), the degradation efficiency of RhB decreased from 100% to 85%, indicating that the oxidative removal of As(III) and the degradation of RhB were competitive reactions. In addition, when the concentration of As(III) was 50 μM and 100 μM, the $FeS_2$/Light/PMS system could reach 100% RhB. Meanwhile, the removal efficiencies of As(III) were 100% and 97.2%, with oxidation efficiencies of 87.3% and 44.9%, respectively. As shown in Fig. 5d-f, when As(III) concentration was maintained at 100 μM and the concentration of RhB was 10, 20 and 30 mg $L^{-1}$, respectively, the FeS$_2$/Light/PMS system achieved complete removal of RhB. Albeit the oxidation rate of As(III) was less than 60%, the removal rates of As(III) reached 100.0%, 97.2% and 68.3%, respectively. These results suggested that the FeS$_2$/Light/PMS system can efficiently remove RhB and As(III) all at once.

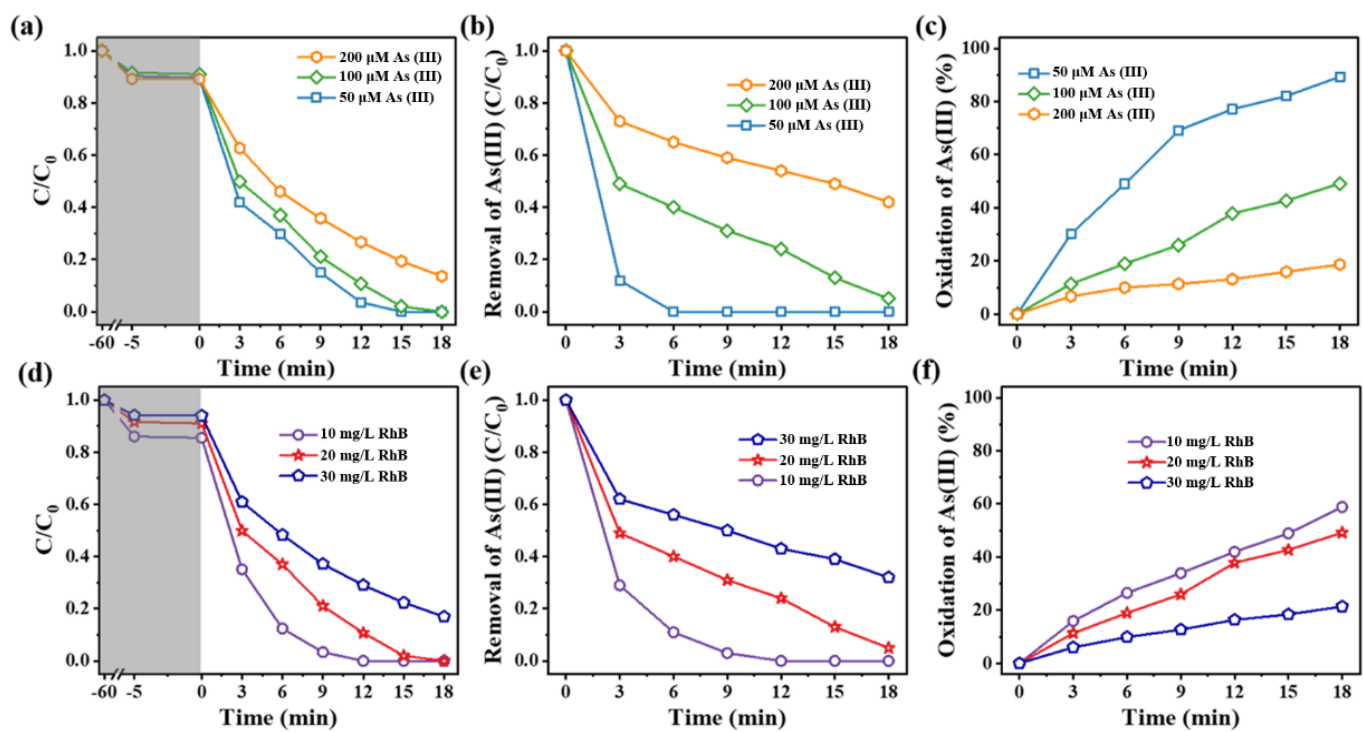

**Fig. 5.** Changes in RhB degradation efficiency (a, d), As(III) removal efficiency (b, e), and As(III) oxidation efficiency (c, f) under different initial concentrations of As(III) (50-200 μM) and 20 mg L$^{-1}$ RhB, as well as different initial concentrations of RhB (10-30 mg L$^{-1}$) and 100 μM As(III) in the FeS$_2$/Light/PMS system. (Reaction conditions: [PMS] = 1.2 mM, [Catalyst] = 0.2 g L$^{-1}$).

*3.3. Reaction mechanisms*

To disclose the degradation mechanism of the FeS$_2$/Light/PMS system, various quenchers were used to identify the dominant ROS in the reaction system [26]. As shown in Fig. 6a, RhB degradation was intensely inhibited after adding 10 mM MeOH, suggesting ·OH and SO$_4$·$^-$ may play a leading role in the reaction process. However, the degradation efficiency only decreased to 90% after adding 10 mM TBA, which could solely scavenge ·OH. The results implied that SO$_4$·$^-$ would be the dominant ROS and the effect of ·OH was limited. PBQ was also applied to identify the existence of ·O$_2^-$, and RhB degradation efficiency decreased by 15% after adding PBQ, indicating that $·O_2^-$ existed but was not the major ROS in the reaction system. On the other hand, $^1O_2$ was also found to be an important ROS in the reaction process. RhB degradation efficiency decreased to 58% after adding TEMP that scavenged $^1O_2$, indicating the nonradical pathway dominated by $^1O_2$ also contributed to the RhB degradation. EPR analyses were carried out to further confirm the reaction process. As illustrated in Fig. 6b-d, the signals of $·OH$, $SO_4^{·-}$, $·O_2^-$ and $^1O_2$ could be captured in the $FeS_2$/Light/PMS system, confirming the results of quenching experiments. Noticeably, signals of DMPO-·OH and $SO_4^{·-}$ were not detected in the $FeS_2$/Light system, indicating that $·OH$, $SO_4^{·-}$ derived from the activation of PMS. Moreover, minor signals of DMPO-·$O_2^-$ and TEMP-$^1O_2$ were observed in the $FeS_2$/Light systems. In the photocatalytic process, photo-generated electrons could reduce the dissolved oxygen and produce $·O_2^-$, which could further be oxidized into $^1O_2$.

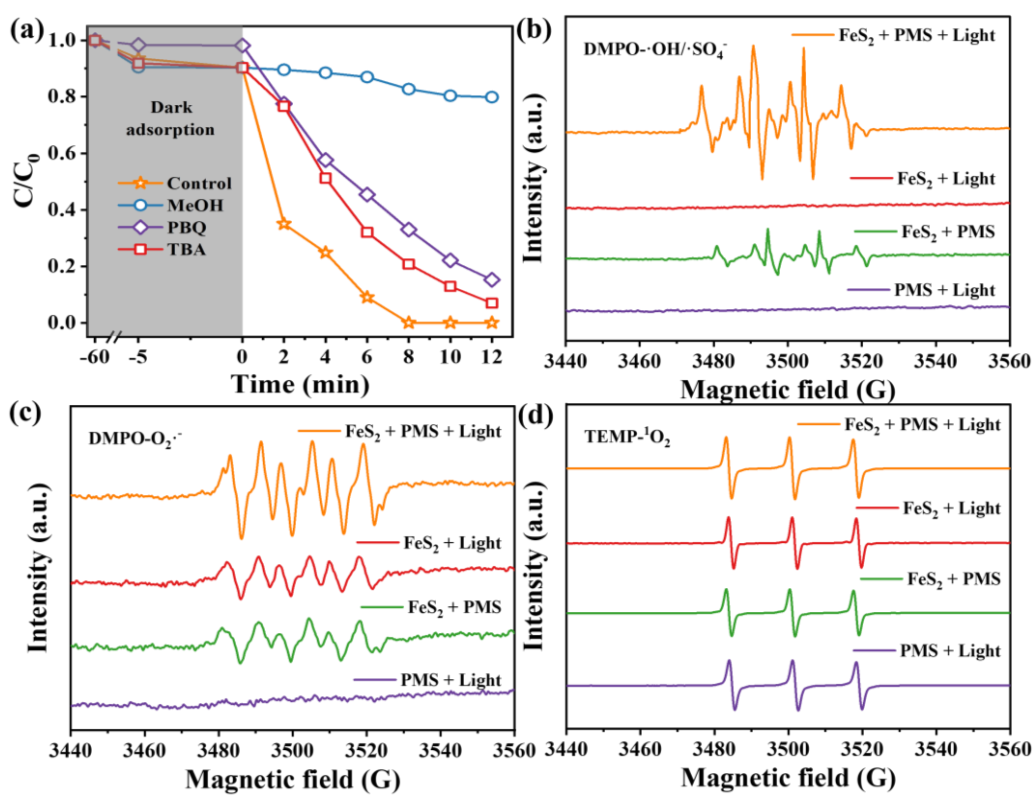

**Fig. 6.** (a) Quenching experiment with different quenchers; EPR spectra of (b) DMPO-·OH/ $SO_4^{·-}$, (c) DMPO-·$O_2^-$ and (d) TEMP-$^1O_2$ in different systems. (Reaction conditions: [PMS] = 1.2 mM, [Catalyst] = 0.2 g L$^{-1}$, [RhB]= 20 mg L$^{-1}$, [quencher] = 10 mM)

To unravel the main reactive centers of the $FeS_2$ catalyst, the $FeS_2$ nanoplate used before and after the reaction were analyzed through XPS (Fig. 7). The proportion of $Fe^{2+}$ on the $FeS_2$ surface decreased from 80.65% to 79.47% after usage, while the proportion of $Fe^{3+}$ increased from 19.35% to 20.53%, as shown in Fig. 7a-b, revealing the existence of redox cycle of $Fe^{2+}/Fe^{3+}$ in the catalytic process [27], therefore, Fe was considered as the main active site of $FeS_2$ catalyst. Furthermore, the $FeS_2$/Light/PMS system showed a decrease in the contents of $S_2^{2-}$ and $S_n^{2-}$ from 60.3% and 20.0% to 55.1% and 19.1%, respectively, as the contents of $S^0$ and $SO_4^{2-}$ increased from 5.1% and 14.7% to 7.0% and 18.9%. This suggested the involvement of $S_2^{2-}$, $S_n^{2-}$, $S^0$, and $SO_4^{2-}$ in the RhB degradation.

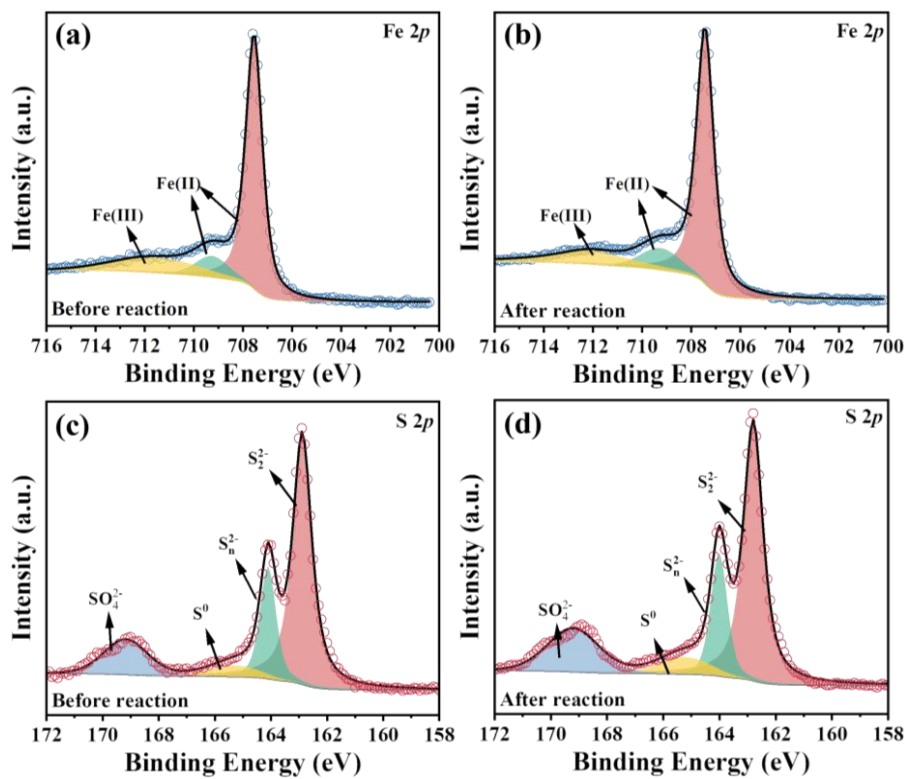

**Fig. 7.** XPS spectra of (a, b) Fe $2p$ and (c, d) S $2p$ in $FeS_2$ before and after the reaction.

Based on the above results, the possible reaction mechanism of the $FeS_2$/Light/PMS system was depicted in Eq. 1-Eq. 9. Under visible light irradiation, $FeS_2$ is firstly excited under illumination to produce photo-generated electron-hole pairs and then $\cdot O_2^-$ as well as $^1O_2$ are produced (Eq. 1) [28]. Meanwhile, $FeS_2$ directly activates PMS to produce $SO_4^{\cdot-}$, $SO_5^{\cdot-}$, $\cdot OH$ and $\cdot O_2^-$ (Eqs. 2-4) [29], while $SO_4^{\cdot-}$ and $SO_5^{\cdot-}$ can further react and convert into $\cdot OH$ and $^1O_2$ (Eqs. 5-6). The $FeS_2$/Light/PMS system has various ROS production pathways, which enable it to efficiently remove RhB and eventually mineralize it into carbon dioxide and water. Moreover, sulfur species such as $S_2^{2-}$ and $S_n^{2-}$ in $FeS_2$ possess strong reduction potential, which can quickly reduce $Fe^{3+}$ to $Fe^{2+}$, thus promoting the cycling of $Fe^{3+}/Fe^{2+}$ in $FeS_2$, improving catalyst stability and generating $S^0$ as well as $SO_4^{2-}$ (Eqs. 7-8).

$$FeS_2 + h\nu \rightarrow e^- + h^+ \rightarrow \cdot O_2^- + {}^1O_2 \tag{1}$$

$$Fe^{2+} + HSO_5^- \rightarrow SO_4^{\cdot-} + Fe^{3+} + OH^- \tag{2}$$

$$Fe^{3+} + HSO_5^- \rightarrow SO_5^{\cdot-} + Fe^{2+} + H_2O \tag{3}$$

$$Fe^{3+} + HSO_5^- + H_2O \rightarrow Fe^{2+} + \cdot O_2^- + SO_4^{2-} + 3H^+ \tag{4}$$

$$SO_4^{\cdot-} + H_2O \rightarrow SO_4^{2-} + \cdot OH + H^+ \tag{5}$$

$$SO_5^{\cdot-} + H_2O \rightarrow {}^1O_2 + 2H^+ + SO_4^{2-} \tag{6}$$

$$Fe^{3+} + S_2^{2-} \rightarrow Fe^{2+} + S_n^{2-} + SO_4^{2-} \tag{7}$$

$$2Fe^{3+} + S_n^{2-} \rightarrow 2Fe^{2+} + S^0 \tag{8}$$

## 4. Conclusions

This work has successfully constructed high-purity pyrite $FeS_2$ nanoplates for the simultaneous removal of dyes and As(III). Various characterizations confirmed the pure pyrite phase in $FeS_2$ and the nanoplate structure, which are critical to exposing more reactive centers for efficient PMS activation in the catalytic removal of dyes and As(III). Under visible light, $FeS_2$ nanoplates could activate PMS efficiently and remove 100% of dye pollutants. It achieved a TOC removal rate of 49.2%, 44.6%, 52.1% and 43.8% for RhB, MB, MG and MO at a concentration of 20 mg L$^{-1}$, while also achieving over 85% As(III) removal efficiency within 18 min. $FeS_2$ nanoplates also maintained excellent stability with low iron leaching. Quenching tests and EPR analysis identified $SO_4^{\cdot-}$ and ${}^1O_2$ as the main ROS in the $FeS_2$/Light/PMS system, revealing the cooperation of radical and non-radical pathways. XPS unveiled the critical roles of sulfur species in retaining the activity and stability of $FeS_2$, for the $S_2^{2-}$ and $S_n^{2-}$ in $FeS_2$ exhibited strong reduction ability that accelerated the redox cycle between $Fe^{2+}$ and $Fe^{3+}$, which enhanced the stability and activity of $FeS_2$. Overall, this work provides a facile approach for preparing high-purity metal sulfide, exhibiting great potential for treating complex dye wastewater.

**Acknowledgments**

This work was supported by the National Key R&D Program of China (2023YFC3900059), the National Natural Science Foundation of China (22078374, 22378434), Key Realm Research and Development Program of Guangdong Province (2020B0202080001), Science and Technology Planning Project of Guangdong Province, China (2021B1212040008), the Scientific and Technological Planning Project of Guangzhou (202206010145).